Deep learning-based flow disaggregation for short-term hydropower plant operations

Duo Zhang[1,*], Luca Petricca[1], Güray Kara[2], Stian Broen[1]

1 Broentech Solutions AS, Langmyra 11b, 3185 Skoppum, Norway

2 Captiva Energi AS, Frøyas gate 15, 0273 Oslo, Norway

* Corresponding author: Duo Zhang, duo@broentech.no

**Abstract**

High temporal resolution data plays a vital role in effective short-term hydropower plant operations. In the majority of the Norwegian hydropower system, inflow data is predominantly collected at daily resolutions through measurement installations. However, for enhanced precision in managerial decision-making within hydropower plants, hydrological data with intraday resolutions, such as hourly data, are often indispensable. To address this gap, time series disaggregation utilizing deep learning emerges as a promising tool. In this study, we propose a deep learning-based time series disaggregation model to derive hourly inflow data from daily inflow data for short-term hydropower plant operations. Our preliminary results demonstrate the applicability of our method, with scope for further improvements.

**Introduction**

The process of refining lower-resolution data into higher resolution is referred to as time series disaggregation (Scher and Peßenteiner, 2021). Proper time series disaggregation can yield numerous benefits for hydropower plant operations. By disaggregating water flow into higher resolutions, reservoir operators can respond more effectively to water releases and storage, thereby ensuring a reliable and stable supply of electricity. The incorporation of disaggregation into hydropower scheduling systems can significantly enhance the precision of energy production planning. Operators can adjust turbine operation schedules more effectively to match fluctuations in electricity demand. The ability of the disaggregation algorithm to provide higher resolution flow patterns is instrumental in improving the evaluation of the ecological effects of hydropower operations on downstream ecosystems and aquatic habitats. Given the increasing variability in weather patterns due to climate change, the application of disaggregation becomes crucial for anticipating extreme events, such as floods or droughts. This, in turn, supports hydropower operators in implementing adaptive strategies.

Theoretically, time series disaggregation refers to the process of estimating the higher temporal resolution time series of a lower temporal resolution time series, aided by one or more higher temporal resolution time series. The lower temporal resolution time series to be disaggregated is often called target series, and the higher temporal resolution time series that helps the process of disaggregation is called indicator series. It's also possible to disaggregate a target series, when an indicator series is not available, while the accuracy of the resulting higher temporal resolution series will be low (Sax and Steiner, 2013).

There are many classical methods for performing time series disaggregation in the field of economy. Such as Denton (Denton, 1971), Denton-Cholette (Dagum and Cholette, 2006), Chow-

Lin (Chow and Lin, 1971), Fernandez (Fernández, 1981) and Litterman (Litterman, 1983).

Denton and Denton-Cholette aim at generating a series that is similar to the indicator series with concern of movement preservation. Chow-Lin, Fernandez and Litterman use one or several indicator series and perform a regression on the low frequency series. A critical assumption of the regression-based methods is that the linear relationship between the higher temporal resolution time series also holds between the lower temporal resolution time series.

For complex systems, disaggregation employing a machine learning/deep learning model can be a reliable choice (Farboudfam et al., 2019). In recent years, machine learning/deep learning has garnered substantial attention owing to its ability to discern intricate patterns and relationships within data. Several studies have investigated the potential of employing machine learning/deep learning techniques for time series disaggregation.

In cases where hydrological records at the desired temporal resolution are available but have relatively short record lengths, machine learning/deep learning models are trained using historical data with known disaggregated components.

Several studies have employed the K-nearest neighbor (KNN) method for rainfall disaggregation (Alzahrani et al., 2023; Park and Chung, 2020; Uraba et al., 2019; Sharif et al., 2013). For a given set of daily records, the KNN method identifies the most similar rainfall events from records in both resolutions. Similarity among rainfall events is evaluated using features such as duration, intensity, magnitude, etc., along with distance metrics as defined in the KNN method. Subsequently, the values are disaggregated based on these similar rainfall events.

Some studies have also explored the use of artificial neural networks (ANN) for disaggregation (Burian et al., 2000, 2001; Poomalai and Chandrasekaran, 2019; Bhattacharyya and Saha, 2023).

Neural network models learn the mapping between the aggregate series and the disaggregated components using data periods that include both resolutions. The input to the neural network consists of the lower resolution records, while the outputs are the corresponding higher resolution values. Afterward, lower resolution values without corresponding higher resolution counterparts are fed into the trained neural network to predict the disaggregated components and complete the missing parts of the higher resolution data.

While the above-mentioned studies primarily leverage feedforward neural networks for disaggregation, in recent years, with the advancements in deep learning, new studies have emerged that employ innovative neural network architectures. Scher and Peßenteiner (2021) demonstrated the potential of utilizing Generative Adversarial Networks (GANs) for disaggregating rainfall to a higher temporal resolution. The trained GANs have shown the ability to generate higher-resolution data reasonably well, based on the conditional daily sum value. However, it's worth noting that the statistics of many generated samples exhibited some deviation from real-world statistics.

In situations where higher temporal resolution records are unavailable, disaggregation can be achieved by constructing a model that leverages the relationship between the target series and the indicator series. Farboudfam et al. (2019) conducted a study on the disaggregation of rainfall time series using artificial neural networks (ANN) and wavelet decomposition. This investigation involved six rain gauges within a catchment area. The indicator series comprised daily data from four of the rain gauges and monthly data from all the rain gauges, while the target series for disaggregation consisted of monthly data from the other two rain gauges. The results indicated that the wavelet + ANN disaggregation model exhibited higher accuracy compared to both ANN and multiple linear regression.

**Methodology and case study**

Operators often require precise inflow forecasts to enhance the accuracy of hydropower plant management, particularly for short-term operations and run-of-river hydropower plants (Yousefi et al., 2023). Considering that the majority of observed hydrological data in Norway are available only at a daily resolution (Vormoor and Skaugen, 2013), there exists a conspicuous gap that needs to be addressed to meet the demands of the Norwegian hydropower industry.

Literature reviews have shown that there are relatively few studies on hydrological time series disaggregation, and the utilization of deep learning for the disaggregation of hydrological data remains relatively unexplored. Many studies are centered around time series data with shorter time spans at both resolutions. These limited time spans may inadequately capture real-world scenarios that encompass longer durations. In some cases, studies rely on nearby observations with both resolutions available as the indicator series, although this practice is infrequently encountered.

With the objective of enhancing short-term hydropower plant operations in Norway, we conducted a case study to explore the potential of employing deep learning for the disaggregation of the most common daily resolution flow records into hourly resolution data. This investigation was carried out under the following prerequisites: 1) The target series are exclusively available at a lower resolution for the entire time span, 2) lower resolution data are the sole data source for all stations, and 3) clear indicator series are not readily available, necessitating their identification.

In the case study, the target time series comprises once-per-day resolution inflow data (daily average inflow data). The objective of the disaggregation process is to generate once-per-hour resolution flow data based on the lower resolution once-per-day data. Considering the rainfall-runoff relationship, weather data available in both once-per-day and once-per-hour resolutions such as rainfall and temperature are selected as the indicator time series.

The inflow data is from The Norwegian Water Resources and Energy Directorate (NVE) Hydrological API (HydAPI, https://hydapi.nve.no), which offers access to NVE's repository of historical and currently observed time series. The series includes the most common hydrological parameters like inflow, water stage, water temperature etc. The weather data for this study is obtained from Open-Meteo (https://open-meteo.com/), an open-source weather API.

To validate the disaggregated results, we selected the Kirkevoll bru flow station, which provides records in both daily and hourly resolutions, for our case study. It is important to note that only the daily resolution records were utilized for model building; the hourly resolution records were exclusively employed for validation purposes. This flow station is located in Vestfold og Telemark county, Norway, with coordinates at latitude 59.69003 and longitude 9.03762. The inflow data spans from December 4, 2018, to January 7, 2021. Weather data was retrieved from the Open-Meteo API using the latitude and longitude coordinates of the flow station.

The foundational mathematical principles underlying time series disaggregation involve the establishment of a statistical relationship between the indicator series and the target series, taking into account fundamental components of the series, such as trend and seasonality, etc., and incorporate additional components like trend variations, irregularities, and other factors that could influence the disaggregation process. In the context of classical disaggregation techniques, we developed a model aimed at learning the general trends from daily resolution time series data, as well as capturing hourly variations from the hourly resolution time series.

The model we designed is illustrated in Figure 1:

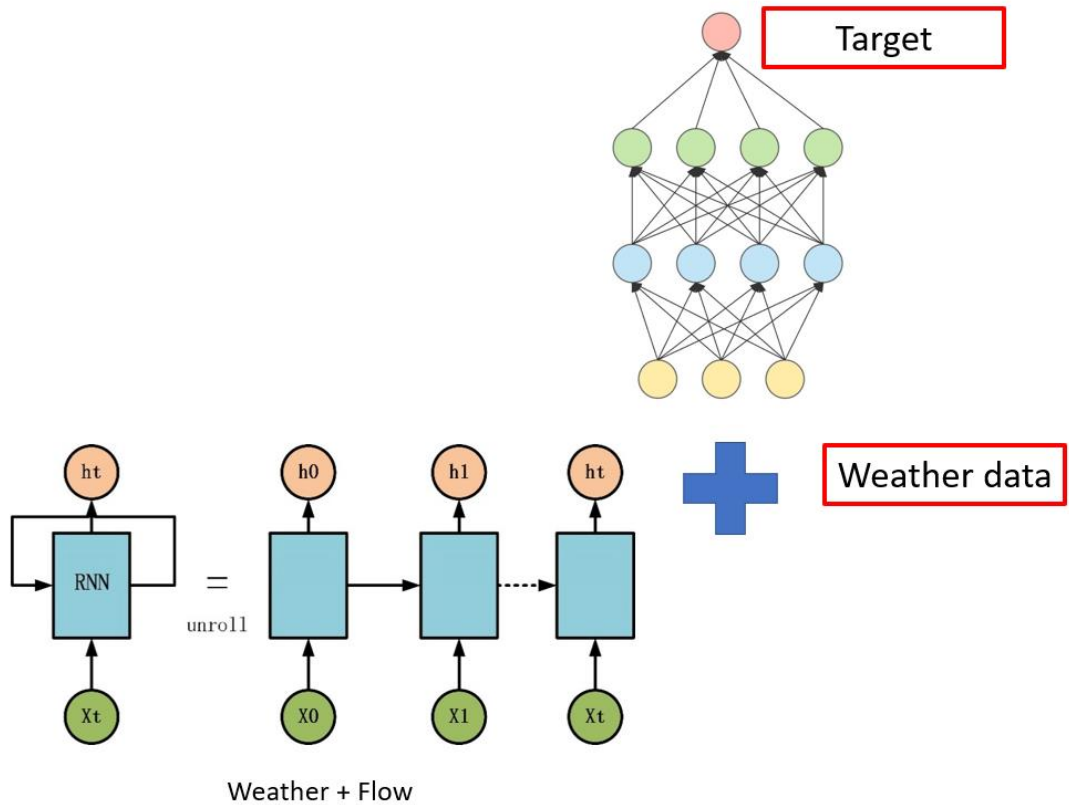

Fig. 1 The flow disaggregation model

The model comprises a combination of a Long Short-Term Memory (LSTM) neural network and a feedforward neural network. The LSTM is employed to process daily resolution weather and flow data from the preceding six days, capturing temporal dependencies and long-term patterns inherent in the data. The LSTM's output is then concatenated with the weather data for the day to be disaggregated, producing the disaggregated flow.

The model operates through three distinct phases: the training phase, prediction phase, and post-processing phase. During the training phase, the LSTM's output, in conjunction with the daily weather data for the target day, is used to predict the daily flow data for that specific day. The loss, representing the difference between the actual values and the predicted values, is employed to train the neural network. Additionally, the LSTM output is combined with each of the hourly weather datasets to feed into the feedforward component, resulting in the generation of 24 predicted flow values. The average of these 24 flow values is then compared with the daily resolution flow data, serving as an additional loss function for neural network training.

After the model has been trained using daily resolution data, the LSTM's output is combined with hourly resolution weather data to make predictions at the hourly resolution level. Finally, as part of the post-processing step, the differences between the average daily flow and the current daily average of the predicted flow values are computed. These differences are then added to each of the 24 predicted values to align their mean with the average daily flow.

The complete algorithm is presented as follows:

Algorithm for training the disaggregation model

**Training phase**

Data: $weather_{day-i}$ : Weather data at day i;

$flow_{day-i}$ : Flow data at day i;

$weather_{hour-i}$ : Weather data at hour i;

$flow_{hour-i}$ : Flow data at hour i;

*# The previous 6 days weather data and flow data is processed by the LSTM*

lstm_output = LSTM ($weather_{day-1} + flow_{day-1}$ , $weather_{day-2} + flow_{day-2}$ , … , $weather_{day-6} + flow_{day-6}$)

*# The LSTM output is combined with the daily weather data at the 7th day, i.e. the day to be disaggregated, and make a prediction through the feedforward neural network*

daily_output = feedforward_nn (lstm_output + $weather_{day-7}$)

*# The loss of daily_output and average flow at the day to be disaggregated is defined as Loss 1*

loss 1 = Loss (daily_output, $flow_{day-7}$)

*# each of the hourly weather data at the day to be disaggregated is combined with the LSTM output, to make a prediction*

For i in range (24):

    $hourly\_output_{hour-i}$ = Feedforward (lstm_output + $weather_{hour-i}$)

*# The mean of the 24 hourly_output and average flow at the day to be disaggregated is used to calculate Loss 2*

hourly_output_mean = ( $hourly\_output_{hour-1}$ + $hourly\_output_{hour-2}$ + ... + $hourly\_output_{hour-24}$) / 24

loss 2 = Loss (hourly_output_mean, $flow_{day-7}$)

*# Use both loss 1 and loss 2 to update the neural networks*

total_loss = loss 1 + loss 2

updated_model = total_loss

**Prediction phase:**

*# To predict flow at hour i, the weather data at hour i at the day to be disaggregated is combined with the LSTM output*

$hourly\_output_{hour-i}$ = Feedforward (lstm_output + $weather_{hour-i}$)

**Post-process:**

*# the difference between the average daily flow and the current daily average of the predicted flow values is added to predicted values*

difference = hourly_output_mean - $flow_{day-7}$

$hourly\_output_{hour-i}$ = $hourly\_output_{hour-i}$ + difference

---

**Result**

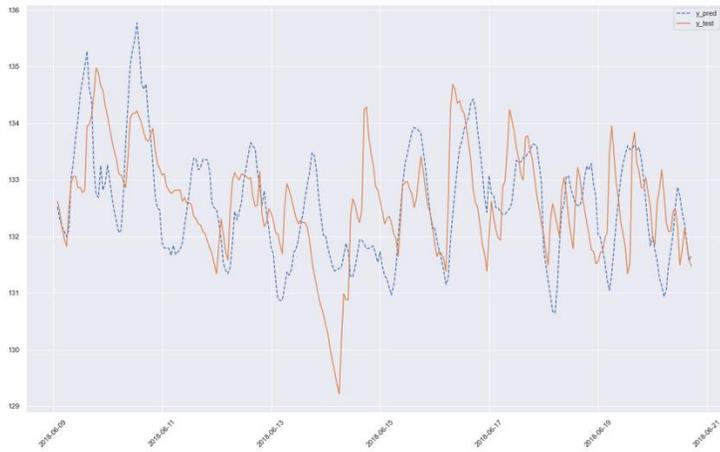
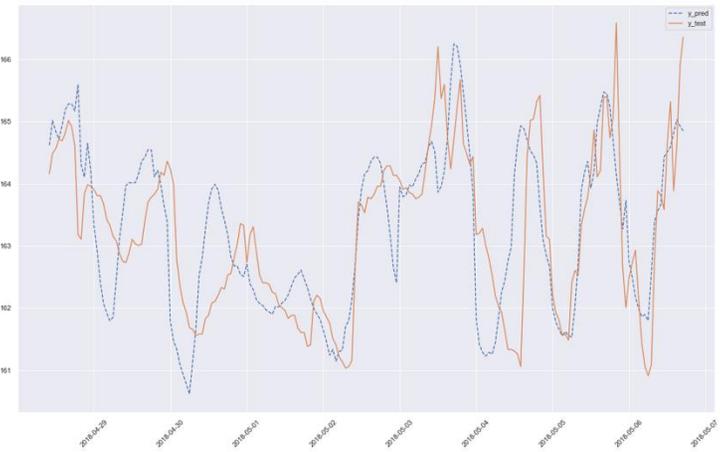

a) b)

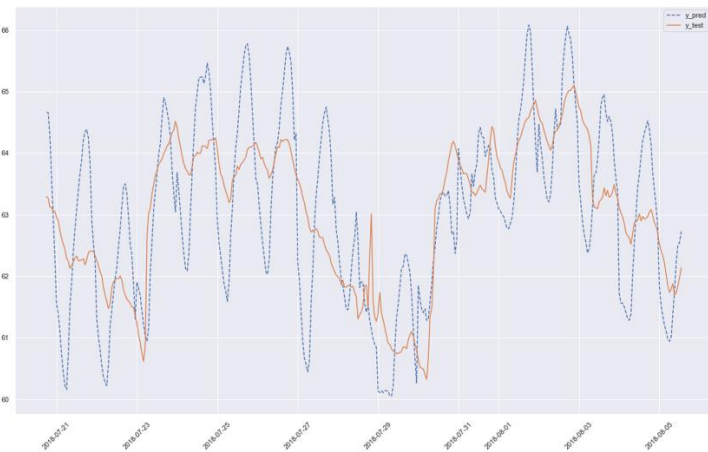

c)

Fig. 2 Result of the disaggregation model

Figure 2 illustrates partial results of the disaggregated hourly values compared to the actual hourly values. The figures reveal that the model successfully generates hourly resolution flow data. However, upon closer examination of specific details, certain weaknesses in the model become apparent. Firstly, the model inaccurately estimates the peak inflow within a day. Secondly, it either overestimates or underestimates certain periodic variations occurring within a day. These issues may be attributed to two potential factors:

1) The weather data, or the indicator time series, may not provide sufficient information to capture variations at a higher temporal resolution effectively.

2) The time lag effect between the hourly time series may not have been adequately studied. For instance, rainfall at the current time may not have an immediate impact on flow; instead, it may exhibit a lagged effect, manifesting itself over hours or even days.

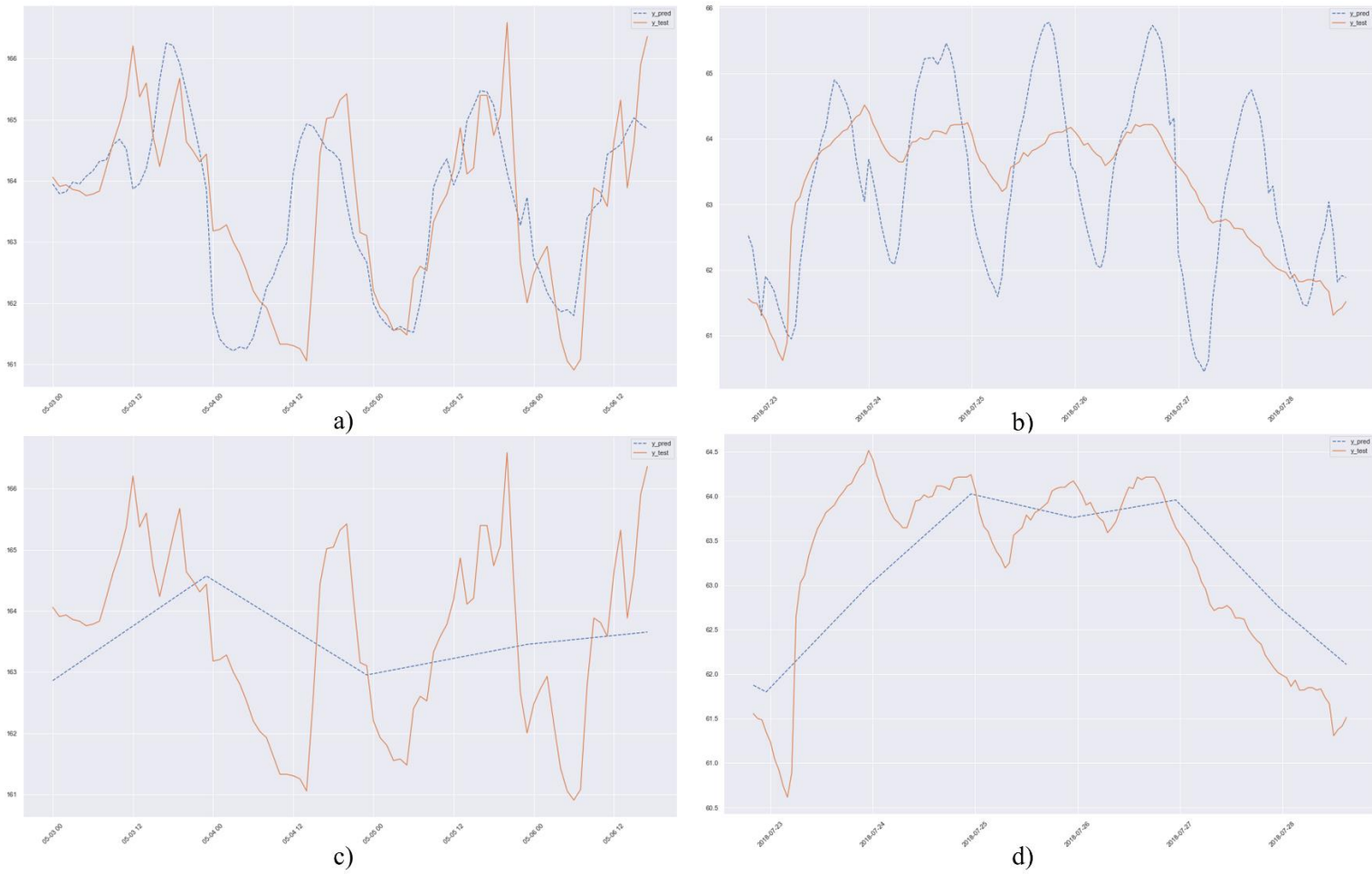

Fig. 3 Comparison between the disaggregation model and linear interpolation

Through visual inspections, we conducted a comparison between the disaggregation model and linear interpolation. Fig 3 a) and Fig 3 b) are disaggregated hourly flow versus the real flow for two scenarios, while Fig 3 c) and Fig 3 d) are linearly interpolated hourly flow versus the real flow, for the corresponding scenarios, respectively. In both scenarios, our model demonstrated the ability to replicate sub-daily variations when compared to linear interpolation. In the second scenario, focusing solely on metrics such as mean absolute error, linear interpolation may initially appear to outperform the disaggregation model. However, upon closer examination of the y-axis (i.e., the flow value), it becomes evident that the error associated with the disaggregation model is smaller than that of linear interpolation, particularly in terms of capturing sub-daily variations. Consequently, the results from the disaggregation model suggest certain advantages over linear interpolation.

**Discussion**

Through a case study, we conducted a preliminary evaluation of the disaggregation model we proposed, relying on visual inspection. While we acknowledge that there are remaining weaknesses in the proposed model, our research underscores the potential of utilizing emerging AI techniques for time series disaggregation, opening avenues for new research. In practical applications, inflow measurements typically yield aggregated data once per day, except for certain large cascaded hydropower plants. Our objective is to provide an algorithm and related application cases for disaggregating daily inflow data into hourly resolution, particularly for short-term hydropower plant operations. Run-of-river hydropower plants, which lack reservoir capacity, stand to benefit significantly from this algorithm, enhancing their operational efficiency.

There is room for enhancing the algorithm's precision in several ways:

1) The quality of the indicator series plays a vital role in time series disaggregation. Improved

quality weather data or potentially other datasets could better capture sub-daily flow variations and enhance model performance.

2) Incorporating time series pre-processing or feature extraction techniques may prove beneficial. Techniques like wavelet transformation for decomposing inflow time series or investigating time step lags using cross-correlation or dynamic time warping could enhance the model's capabilities.

3) Exploring the use of an attention mechanism as a supplement to the LSTM component could be advantageous. The attention mechanism considers outputs from all time steps, as opposed to using only the last output as in the current model. This could provide the model with a better understanding of the time series.

Integrating the algorithm into real-world applications also presents some challenges. One of the primary challenges in implementing the algorithm within the load forecasting system for multiple hydropower plants is the necessity for individualized optimization and tuning for each facility. While the algorithm demonstrates strong performance in initial testing environments, its effectiveness may vary across different hydropower installations due to the distinct characteristics of water flow dynamics and landscape geology. Specialized calibration and fine-tuning, particularly in the selection of indicator time series, are imperative to adapt the algorithm to each hydropower facility, ensuring accurate load forecasts tailored to their specific operational context.

**Acknowledgement**

We gratefully acknowledge Broentech Solutions AS, which has supported this study and is currently in the process of implementing the algorithm into the Tyde™ forecasting system. We also extend our appreciation to Captiva Asset Management AS, a leading company providing management services for renewable assets owners, for their invaluable contributions and initial ideas that greatly enriched this research.

1533-1550.

Yousefi, M., Wang, J., Fandrem Høivik, Ø., Rajasekharan, J., Hubert Wierling, A., Farahmand, H., & Arghandeh, R. (2023). Short-term inflow forecasting in a dam-regulated river in Southwest Norway using causal variational mode decomposition. Scientific Reports, 13(1), 7016.

Uraba, M. B., Gunawardhana, L. N., Al-Rawas, G. A., & Baawain, M. S. (2019). A downscaling-disaggregation approach for developing IDF curves in arid regions. *Environmental monitoring and assessment*, *191*, 1-17.

Vormoor, K., & Skaugen, T. (2013). Temporal disaggregation of daily temperature and precipitation grid data for Norway. *Journal of Hydrometeorology*, *14*(3), 989-999.